# Definition of Strange Attractor in Benard Problem for Generalized Couette Cell

V.V.Gotsulenko, L.A.Gaponova, P.I.Kogut
"Strategy" the Institute for Entrepreneurship,
Zhovti Vody, Ukraine

ABSTRACT. For movements of the viscous continuous flow in generalized Couette cell the dynamic system describing the central limiting variety is received.

Couette cell is traditional modeling object for studying of many problems of hydrodynamics. Properties of liquid flow and its bifurcation with growth of Reynolds number in Couette cell with several internal cylinders actually are not studied. With growth of internal cylinders number computer modeling of a liquid flow in Couette cell becomes impossible. It is caused, in particular by the fact that since on each cylinder the separate limiting condition at any digitization of the initial equations of mathematical model (a method of final differences, a method of final elements) we come to system of the algebraic equations which quantity is so huge that their decision becomes problematic even on modern computers. The purpose of the given work is obtaining at $\varepsilon \to 0$ (the parameter $\varepsilon$ characterizes diameters of internal cores) and the analysis of the concentrated dynamic system describing the central limiting variety for movements of the viscous poorly compressed continuous flow in generalized Couette cell.

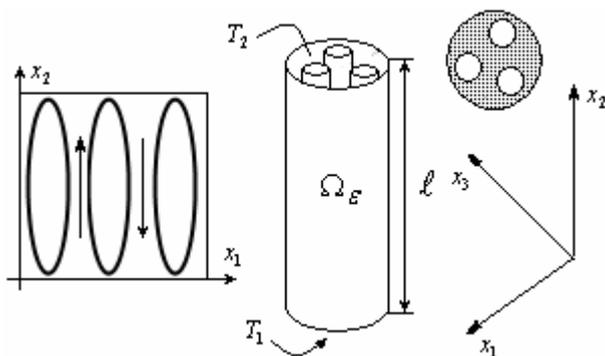

**Figure 1.** The scheme of generalized Couette cell $\Omega_\varepsilon$

As shown in works [KGL08-GL07], limiting characteristics of the continuous flow at $\varepsilon \to 0$ satisfy to following system of Birkman equations on set $\Omega$:

$$\begin{cases} \partial V/\partial t + (V\nabla)V + (2\pi/C_0)(V - V^*) = -(1/\rho)grad(p) + \nu\Delta V + \overline{g}, \\ \partial\rho/\partial t + div(\rho V) = 0, \\ \partial T/\partial t + (2\pi/C_0)(T - T^*) + div(T \cdot V) = k\Delta T, \\ \rho = \rho_0(1 - \gamma(T - T_2)), \end{cases} \quad [1]$$

where limiting parameters $V^*$ and $T^*$ are connected with limiting transition structure and value of a constant $C_0$ [KGL08-GL07].

Let's note, that by virtue of cylindrical symmetry it is possible to pass from 3-dimensional statement of an initial problem to 2-dimentional, considering section of a cell of generalized Couette cell along an axis of the external cylinder. Then all variables will not depend from $x_3$, and $x_3$ - a component of speed will be absent. Further we use Bussinesc approach which, as we know, consists that the liquid is supposed poorly compressed and dependence of density on temperature is considered in the equations only in one place, in the right part of the equation for speed. We shall enter also [K01] replacement of variables:

$$p = p_0 - \rho_0 g[1 - \gamma(T - T_2)]x_2 + \tilde{p}(x_1, x_2, t) \quad T = T_0 + \frac{T_1 - T_2}{\ell}x_2 + \vartheta(x_1, x_2, t),$$

where $\tilde{p}(x_1, x_2, t)$ - a deviation of a field of pressure from hydrostatic pressure $p_0 - \rho g x_2$, and $\vartheta(x_1, x_2, t)$ - a deviation of temperature from a linear structure, and we use in the right part of the first equation [1] following representation [K01]:

$$\frac{1}{\rho} grad(p) = \frac{-\rho_0 g\bar{j} + grad(\tilde{p})}{\rho_0 g[1-\gamma(T-T_2)]} \approx -g\bar{j} - g\gamma(T-T_2)\bar{j} + \frac{1}{\rho_0} grad(p). \quad [2]$$

Considering, that $\bar{g} = -\bar{j}g$ we copy the equations [1] in the form of:

$$\begin{cases} \partial V/\partial t + (V\nabla)V + (2\pi/C_0)(V-V^*) = \gamma g \vartheta \bar{j} - (1/\rho_0) grad(p) + \nu \Delta V, \quad div(V)=0, \\ \partial \vartheta/\partial t + (2\pi/C_0)(\vartheta-\vartheta^*) + div(\vartheta \cdot V) - \frac{T_1-T_2}{\ell} div(x_2 V) = k \Delta \vartheta. \end{cases} \quad [3]$$

On the top and bottom of Couette cell it is imposed the boundary conditions expressing a constancy of temperature and absence of a stream of a liquid through border:

$$\vartheta\big|_{x_2=0} = 0 \quad \vartheta\big|_{x_2=\ell} = 0 \quad (V \cdot \bar{j})\big|_{x_2=0} = 0 \quad (V \cdot \bar{j})\big|_{x_2=\ell} = 0. \quad [4]$$

Let's present the vector equations [3] in the scalar form, designating accordingly through $x_1$ and $x_2$ - components of speed $V$ through $u$ and $\upsilon$:

$$\begin{cases} \partial u/\partial t + u\, \partial u/\partial x_1 + \upsilon\, \partial u/\partial x_2 + (2\pi/C_0)(u-u^*) = -\rho_0^{-1} \partial \tilde{p}/\partial x_1 + \nu(\partial^2 u/\partial x_1^2 + \partial^2 u/\partial x_2^2) \\ \partial \upsilon/\partial t + u\, \partial \upsilon/\partial x_1 + \upsilon\, \partial \upsilon/\partial x_2 + (2\pi/C_0)(\upsilon-\upsilon^*) = \gamma g \vartheta - \rho_0^{-1} \partial \tilde{p}/\partial x_1 + \nu(\partial^2 \upsilon/\partial x_1^2 + \partial^2 \upsilon/\partial x_2^2), \\ \partial \vartheta/\partial t + \partial(\vartheta u)/\partial x_1 + \partial(\vartheta \upsilon)/\partial x_2 + (2\pi/C_0)(\vartheta-\vartheta^*) - \frac{T_1-T_2}{\ell}\upsilon = k(\partial^2\vartheta/\partial x_1^2 + \partial^2\vartheta/\partial x_2^2), \\ \partial u/\partial x_1 + \partial \upsilon/\partial x_2 = 0. \end{cases} \quad [5]$$

Let's exclude a field of pressure by differentiation of the first equation on $x_2$, and the second on $x_1$ and then we subtract one of another. As a result we receive the equation:

$$(\partial/\partial t)(\partial u/\partial x_2 - \partial \upsilon/\partial x_1) = -u\, \partial^2 u/\partial x_1 \partial x_2 - \upsilon\, \partial^2 u/\partial x_2^2 + u\, \partial^2 \upsilon/\partial x_1^2 + \\ + \upsilon\, \partial^2 \upsilon/\partial x_1 \partial x_2 - \gamma g\, \partial \vartheta/\partial x_1 + \nu(\partial^3 u/\partial x_1^2 \partial x_2 + \partial^3 u/\partial x_2^3 - \partial^3 \upsilon/\partial x_1 \partial x_2^2 - \partial^3 \upsilon/\partial x_2^3) \quad [6]$$

Further, we apply Galerkin method, representing required fields in the form of decomposition in rows on some full system of basic functions; after that dependence on time of decomposition factors becomes a subject of consideration. We shall build decomposition on basis of trigonometrically functions of a kind:

$$\sin(m\alpha x_1)\sin(n\beta x_2) \quad \sin(m\alpha x_1)\cos(n\beta x_2) \quad \cos(m\alpha x_1)\sin(n\beta x_2)$$

where $n$ and $m$ - integers, and $\alpha, \beta$ - constants connected with geometry of Benard convective shafts [K01]. So, we suppose

$$\vartheta(x_1, x_2, t) = \sum_{m=0}^{\infty} \sum_{n=1}^{\infty} F_{mn}(t) \cos(m\alpha x_1) \sin(n\beta x_2) \qquad [7]$$

From a condition of zero divergence $\partial u/\partial x_1 + \partial v/\partial x_2 = 0$ follows, that there is a function of a flow

$$\psi(x_1, x_2, t): \quad u = -\partial\psi/\partial x_2 \quad v = -\partial\psi/\partial x_1$$

decomposing which in a trigonometrically row on the chosen basis, we receive:

$$\psi(x_1, x_2, t) = \sum_{m=1}^{\infty} \sum_{n=1}^{\infty} G_{mn}(t) \sin(m\alpha x_1) \sin(n\beta x_2). \qquad [8]$$

Then for a component of speed we have

$$\begin{cases} u(x_1, x_2, t) = -\sum_{m=1}^{\infty} \sum_{n=1}^{\infty} n\beta G_{mn}(t) \sin(m\alpha x_1) \cos(n\beta x_2), \\ v(x_1, x_2, t) = -\sum_{m=1}^{\infty} \sum_{n=1}^{\infty} m\alpha G_{mn}(t) \cos(m\alpha x_1) \sin(n\beta x_2). \end{cases} \qquad [9]$$

Further, substituting expressions [7] and [9] in the equations [6] and, using parities of orthogonality for basic functions we receive infinite system of the equations for factors $G_{mn}$ and $F_{mn}$. By analogy as in work [K01] we shall break the received rows on the first harmonics, considering essential and distinct from zero $G_{11}$, $F_{11}$, $F_{02}$ which we shall designate, accordingly, through $X$, $Y$ and $Z$. So, we suppose:

$$u = -X\beta \sin(\alpha x_1)\cos(\beta x_2) \quad \upsilon = X\alpha \cos(\alpha x_1)\sin(\beta x_2) \quad \vartheta = Y\beta \cos(\alpha x_1)\sin(\beta x_2) - Z\sin(2\beta x_2). \quad [10]$$

Let's substitute these expressions in the first equation [6]. In the received parity all arising combinations of sinuses and cosines it is necessary to result by means of trigonometrically formulas to the sums of elements of the chosen basis, and then to reject elements, different by structure from a unique combination of a kind $\sin(m\alpha x_1)\sin(n\beta x_2)$ present at the left part. Equating factors in the left and right part, we receive

$$\frac{dX}{dt} = \frac{\alpha \gamma g}{\alpha^2 + \beta^2}Y - \left[\nu(\alpha^2 + \beta^2) + \frac{2\pi}{C_0}\right]X. \qquad [11]$$

Acting similarly with the second equation [7], we shall receive two equations:

$$\frac{dY}{dt} = \alpha \frac{T_1 - T_2}{\ell}X - k(\alpha^2 + \beta^2)Y - \alpha\beta X \cdot Z \quad \frac{dZ}{dt} = -4k\beta^2 Z + \frac{\alpha\beta}{2}X \cdot Z. \qquad [12]$$

Doing replacement of variables:

$$X = Ax \quad Y = By \quad Z = Cz, \quad t = D\tau$$

with values of constants $A$ $B$, $C$ and $D$ from [K01], we shall receive, that the dynamic system [10-12] will be transformed to a kind:

$$\frac{dx}{d\tau} = \sigma y - \left[\sigma + \frac{q}{C_0}\right]x \quad \frac{dy}{d\tau} = rx - y - xz \quad \frac{dz}{d\tau} = -bz + xz, \qquad [13]$$

where parameters $\sigma$ $r$, $b$ are defined, for example in [K01] $q = 2\pi/k(\alpha^2 + \beta^2)$. We shall note that at $q = 0$ or $C_0 = \infty$ the system [13] passes in Lorentz's classical system.

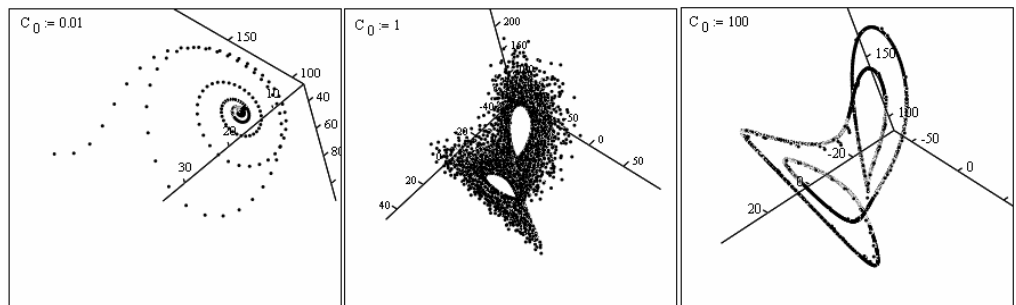

**Figure 2.** Reorganization of system attractor [14] at a variation of parameter $C_0$ at $\sigma = 10$ $b = 8/3$, $r = 27$ and $q = 1$